\documentclass[USenglish,twocolumn]{article}

\usepackage[utf8]{inputenc}%(only for the pdftex engine)
\usepackage[big]{dgruyter}
\usepackage{upgreek}

\begin{document}

  \articletype{Research Article{\hfill}Open Access}

  \author*[1]{Sharina M.E.}

\author[2]{Maricheva M.I.}
%affiliation should be placed in lines starting with "{ \let\thempfn\relax%" after \maketitle (see below)
\affil[1]{Special Astrophysical Observatory, Russian Academy of Sciences, Nizhnii Arkhyz, 369167 Russia, E-mail: sme@sao.ru}
\affil[2]{Special Astrophysical Observatory, Russian Academy of Sciences, Nizhnii Arkhyz, 369167 Russia, E-mail: marichevar@gmail.com}

  \title{\huge Chemical composition and ages of four globular clusters in M31 from the analysis of their integrated-light spectra}

  \runningtitle{Chemical composition and ages of four globular clusters in M31 from the analysis of their integrated-light spectra}

  %\subtitle{...}

  \begin{abstract}
{We compare the results on the chemical composition of four globular clusters (GCs) in M31 (Bol6, Bol20, Bol45, and Bol50) (Maricheva 2021. Study of integrated spectra of four globular clusters in M 31.
 Astrophys. Bull. 76:389–404. doi: https://doi.org/10.1134/S199034132104009X) to  the available literature data on integrated-light spectroscopy of
 globular clusters with similar ages and chemical abundances in our
 Galaxy and M31 and on the chemical abundances of stars in two galaxies. 
The clusters and their literature analogues are of moderate metallicity
 $ \rm -1.1 < [Fe/H] <-0.75 $~dex and older (10 Gyr). 
 Mg, Ca, and C abundances of four GCs are higher than literature estimates for the GCs in M31 with $\rm [Fe/H]\sim -1$~dex obtained using 
high-resolution integrated-light spectroscopy methods.
Using literature data, we did not find complete analogues for the studied clusters in our Galaxy and M31 in terms of age, helium mass fraction
 (Y), and chemical composition. The alpha-element abundances in four clusters are about 0.2 dex higher than the average for stars in the  
Galactic field at $\rm [Fe/H] \sim -1 $~dex. We suggest that these and  Maricheva's (Maricheva M. 2021) findings about lower metallicities of the studied 
objects than the average metallicity of red giants in the M31 halo and about the abundances of alpha-process elements in them corresponding to 
the average value for stars in the M31 inner halo  likely indicate that the star formation process in the vicinity of M31 at the time of our 
sample clusters formation was complex with the inflow of fresh gas from the intergalactic medium and violent star forming events associated 
with SNe~II bursts.}
\end{abstract}
%  \classification[PACS]{}
 % \communicated{...}
 % \dedication{...}

  \journalname{Open Astronomy}
\DOI{DOI}
  \startpage{1}
  \received{..}
  \revised{..}
  \accepted{..}

  \journalyear{2014}
  \journalvolume{1}
%  \journalissue{1}

\maketitle
%\affil[1]{Special Astrophysical Observatory, Russian Academy of Sciences, Nizhnii Arkhyz, 369167 Russia, E-mail: marichevar@gmail.com}
%\affil[2]{Special Astrophysical Observatory, Russian Academy of Sciences, Nizhnii Arkhyz, 369167 Russia, E-mail: ////}

%lines connected to author's affiliation

%{ \let\thempfn\relax% Remove footnote number printing mechanism
%\footnotetext{\hspace{-1ex}{\Authfont\small \textbf{Corresponding Author: Sharina M.E.:}} {\Affilfont Special Astrophysical Observatory, Russian Academy of Sciences, %Nizhnii Arkhyz, 369167 Russia; Email: sme@sao.ru}}
%}

%{ \let\thempfn\relax% Remove footnote number printing mechanism
%\footnotetext{\hspace{-1ex}{\Authfont\small \textbf{Maricheva M.I.:}} {\Affilfont Special Astrophysical Observatory, Russian Academy of Sciences, Nizhnii Arkhyz, 369167 Russia}}
%}

\section{Introduction}

The globular cluster (GC) system in the nearest to our Galaxy giant spiral, M31, has been studied in detail in the literature. However, in order to determine ages 
and metallicities of GCs, their colours and spectral indices were commonly used in comparison with models of stellar populations. For several objects, 
integrated-light spectroscopic studies were performed in order to determine their chemical composition (\citealt{Col14} [hereafter: C14],
 \citealt{Sak16} [hereafter: S16], \citealt{Sh18}, \citealt{L18}) and age (C14). To date, the deepest colour-magnitude diagrams (CMDs) for clusters in M31 
reach confidently the level of the horizontal branch (\citealt{Fe12}, \citealt{L21}).

In this article, we consider the results of the analysis by \cite{Mar21}(hereafter: MMI21) of integrated-light (hereafter: IL) spectra of four GCs in M31. 
Our aim is the comparison of the determined  ages, helium mass fraction (Y), and the abundances of chemical elements Fe, C, N, Mg, Ca, and Ti with the 
corresponding data for Galactic and M31 GCs and with the data for field stars in two galaxies. The comparison of the data for GCs with similar ages and 
metallicities in different galaxies and different galactic subsystems will ultimately help us to answer the question: how environment and star formation 
history influence the chemical composition of GCs?

\section{Spectroscopic data, methods of their analysis and main results}

The spectra of four GCs were obtained in 2020 with the 6-m SAO RAS telescope using the SCORPIO-1 multimode focal reducer (\citealt{Af05}) in the long slit mode.
 The used grism VPHG1200B and the long slit of 1 arcsec provide the resolution FWHM $\sim 5.5$~\AA~ (full width at half maximum of a single spectral line)
 within the spectral range 3600-5400~\AA. Please see the paper by MMI21 describing in detail the observational data and the procedure of their reduction 
and analysis. 

Ages, Y, and the abundances of chemical elements in GCs can be determined using their medium-resolution IL spectra (e.g. \citealt{Sh20} and references therein). 
For this, the observed spectra are approximated by the synthetic ones, calculated on the basis of models of stellar atmospheres. The parameters of stellar 
atmospheres are set by the selected theoretical isochrone of stellar evolution. The synthetic stellar spectra are summed according to the selected stellar 
mass function. We apply air wavelengths for our analysis\footnote{The IAU standard for conversion from air to vacuum wavelengths is given in the study by \cite{Morton91}.}. 
The method was tested with the medium-resolution spectra from the \cite{Sch05} library (\citealt{Sh20}). The method is implemented in the program {\it CLUSTER} 
(\citealt{Sh20} and references therein) based on the plane-parallel hydrostatic models of stellar atmospheres calculated with the ATLAS~9 code (\citealt{CK03}).
 A proper isochrone is selected by fitting the lines of the Balmer series of hydrogen and CaI~4227,
and K, and H CaII~3933.7, and 3968.5~\AA\ lines by the model ones. Following the previous studies, MMI21 applied the stellar mass function by \cite{Cha05} and 
the scaled-solar isochrones by \cite{B08} for the analysis. The chemical composition of the \cite{B08} stellar evolutionary models is given relative to the solar
 abundances of \cite{Grevesse98}, in the same abundance scale as the \cite{CK03} models.
Additionally, MMI21 used \cite{P04} isochrones. These models were computed relative to a scaled solar heavy-element distribution from \cite{GN93}.
MMI21 did not find any systematic differences between the chemical abundances derived on the basis of the \cite{B08} and \cite{P04} isochrones. 
%The development of isochrones according to the initial chemical composition of the studied cluster  and taking into account the evolution of this chemical composition over time is a future question.} 

 The abundances of the chemical elements for Bol20, Bol50 were derived by MMI21 for the first time. For all four objects (Bol6, Bol20, Bol45, and Bol50), Y 
was determined for the first time. All the studied clusters turned out to be older than 10 Gyr with the metallicities in the range:
 from $\rm-0.75 < [Fe/H] < -1.1$~dex \footnote{The abundance of iron in solar units: [Fe/H]=$\rm log(N_{Fe}/N_{H})-log(N_{Fe}/N_{H})\odot$, 
where  $\rm N_{Fe}/N_{H}$ is the ratio of the concentrations of iron and hydrogen by the number of atoms or by mass. The mass fractions of hydrogen X,
 helium Y, and metals Z for the Sun are given by \cite{As09}. $\rm X+Y+Z=1$.} : $\rm T=11.2\pm 1$~Gyr, $\rm Y=0.3\pm0.05$, and $\rm [Fe/H]=-0.75\pm0.1$~dex 
for Bol~6, $\rm T=13\pm 1$~Gyr, $\rm Y=0.26\pm0.01$, and $\rm [Fe/H]=-1.0\pm0.1$~dex for Bol~20 and $\rm T=11\pm 1$~Gyr, $\rm Y=0.26\pm0.01$, 
and $\rm [Fe/H]=-1.1\pm0.1$~dex for Bol~45 and Bol~50. The spectra and the chemical composition of Bol~45 and Bol~50 appeared to be very similar. 
 A reasonable agreement was found between the chemical abundances for Bol6 and Bol45, obtained by MMI21, 
and the corresponding literature data of high-resolution IL spectroscopy from C14 and S16. The ages of four GCs agree well with the literature ones from 
C14 and \cite{Ca16}.
% (please, see the infrared data in S16). 

\begin{figure*}
\includegraphics[scale=0.42, angle=270]{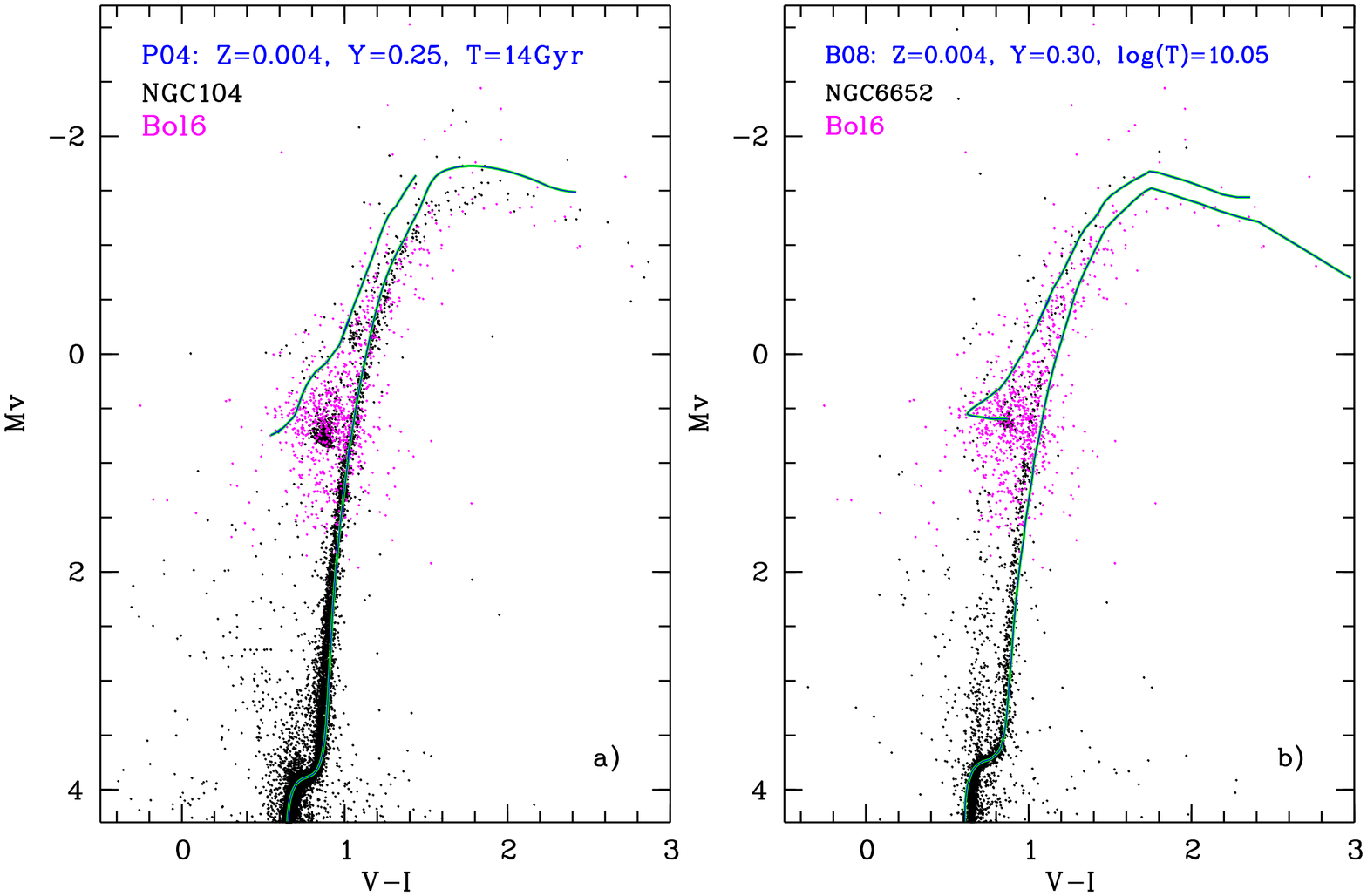}
\includegraphics[scale=0.42, angle=270]{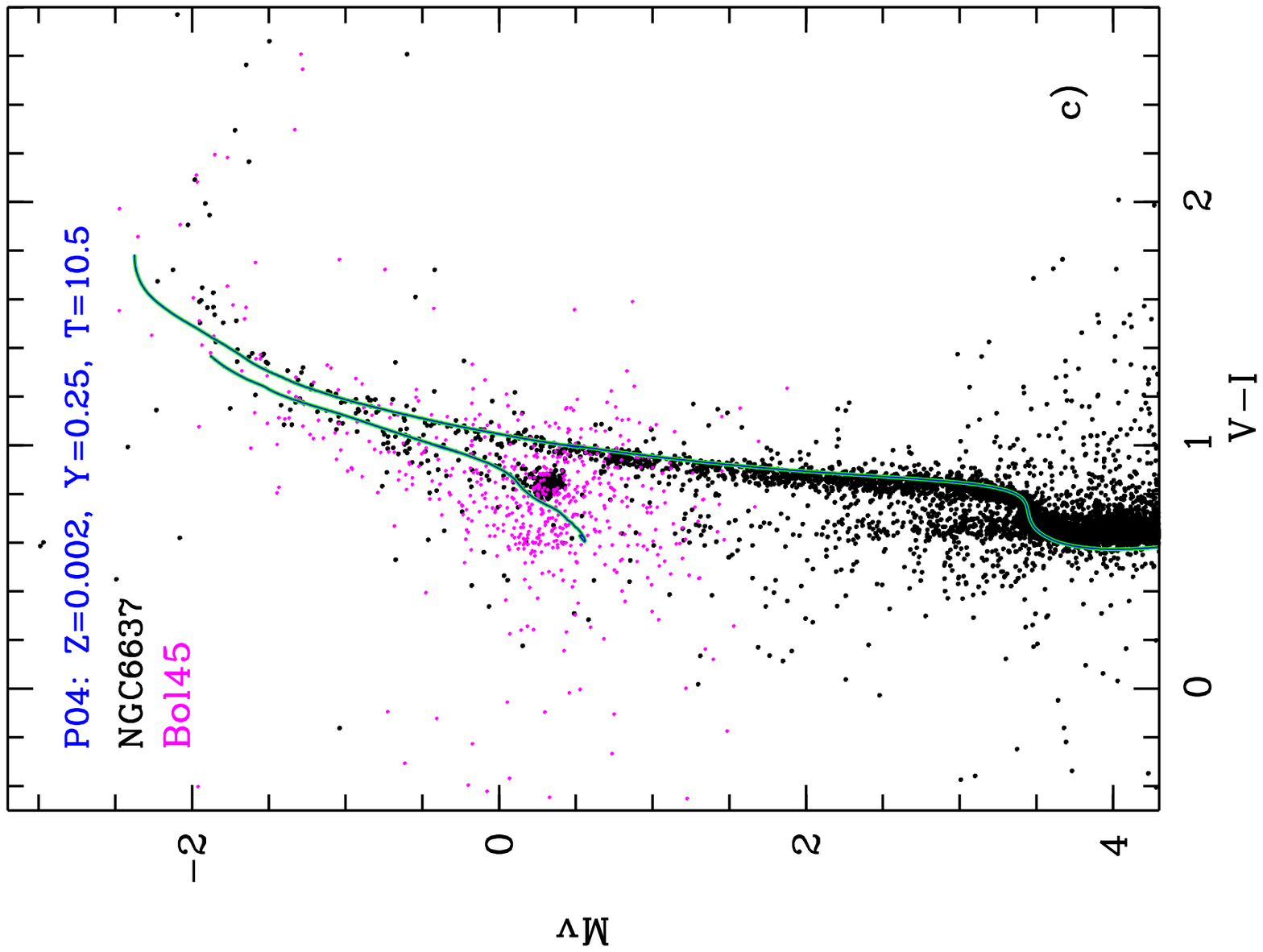}
\caption{CMDs for Bol~6 (panels a and b) and Bol~45 (panel c) (\citealt{Fe12}) (magenta dots) in comparison with the CMDs of Galactic GCs (\citealt{Sar07}) NGC~104, 6652 and 6637 (black dots) and with the evolutionary isochrones (green lines) by \cite{B08} for Bol~6 (panel b) and \cite{P04} for Bol~6 (panel a) and for Bol~45 (panel c), selected by MMI21 for the analysis of the IL spectra of the GCs.}
\label{fig1}
\end{figure*}

\begin{figure*}
\includegraphics[scale=0.6, angle=270]{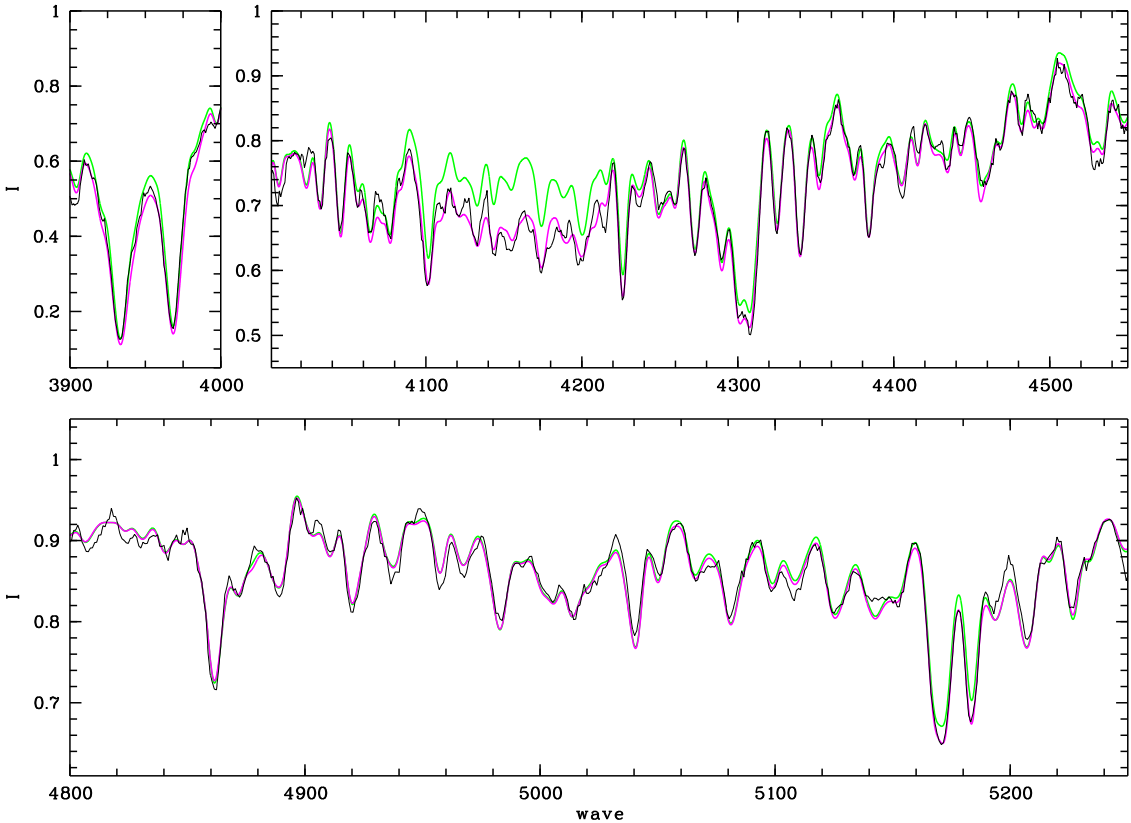} 
\caption{IL spectrum of Bol~6 from MMI21 (in black) in comparison with the synthetic one calculated using the isochrone from \cite{P04} (see Fig. \ref{fig1}, panel a) and the chemical composition defined by MMI21 (magenta). A synthetic spectrum with the modified chemical composition is shown in green with Ca, Mg and C depleted by 0.2 dex and N depleted by 0.5~dex.}
\label{fig2}
\end{figure*}

Fig.~\ref{fig1} demonstrates a comparison between the CMDs of Bol~6 and Bol~45 (\citealt{Fe12}), the CMDs of Galactic GCs from \cite{Sar07}, and the evolutionary
 isochrones  by \cite{B08} and \cite{P04} selected by MMI21 for the analysis of the IL spectra of Bol~6 and Bol~45. MMI21 concluded that there are no Galactic
 GCs in the library of \cite{Sch05} with the spectra very similar to the spectra of four M31 GCs. The Galactic GCs with the CMDs shown in Fig.~\ref{fig1} 
demonstrate that the spectra are most appropriate for the comparison. 
In Fig.~\ref{fig2}, we illustrate the procedure of spectra fitting. The  IL spectrum of Bol~6 analysed by MMI21 is compared with the synthetic one calculated
 using the isochrone (Fig.~\ref{fig1}, panel a) and the chemical composition defined by MMI21 using the method by \cite{Sh20}. The green line indicates the 
synthetic spectrum with the decreased abundances of C, N, Ca, and Mg.

\section{Comparison of the chemical abundances, determined by Maricheva(2021),
with the abundances for GCs and field stars in our Galaxy and M31}

\begin{figure*}
\includegraphics[scale=0.53, angle=270]{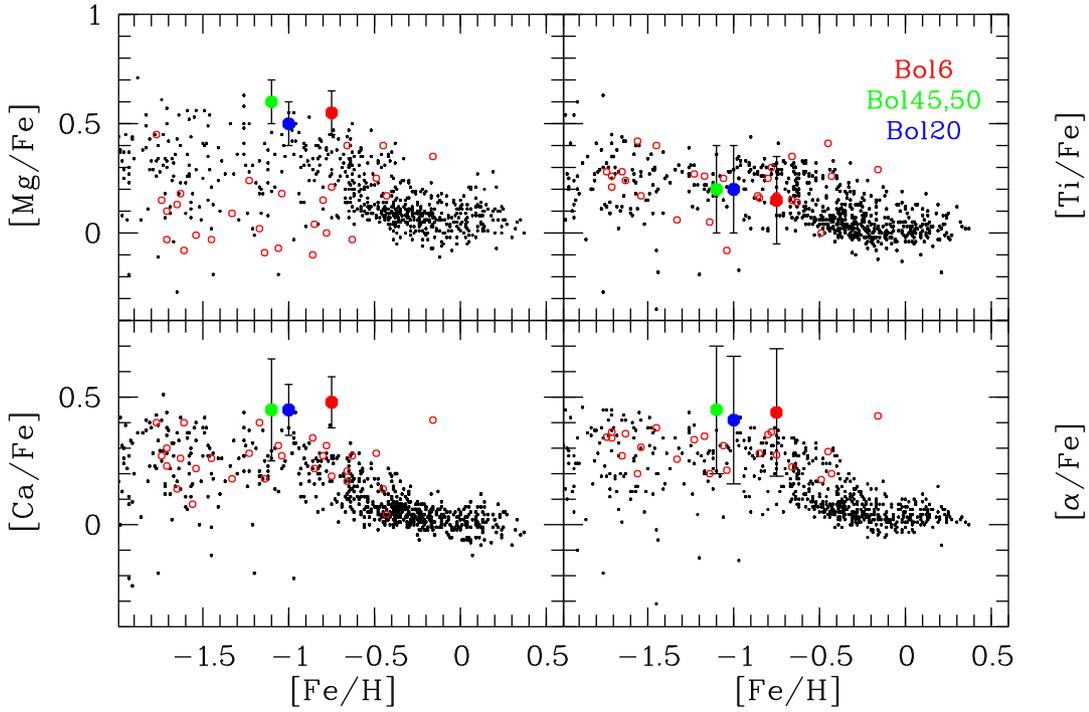} 
\caption{Ca, Mg and Ti abundances and [$\rm \alpha$/Fe] measured by MMI21 for Bol6, Bol20, Bol45, and Bol50 (large filled circles) in comparison with that of Galactic field stars from \cite{Vnn04} (black small dots) and with that of GCs in M31 estimated by C14 (open circles).}
\label{fig3}
\end{figure*}

\begin{figure*}
\includegraphics[scale=0.53, angle=270]{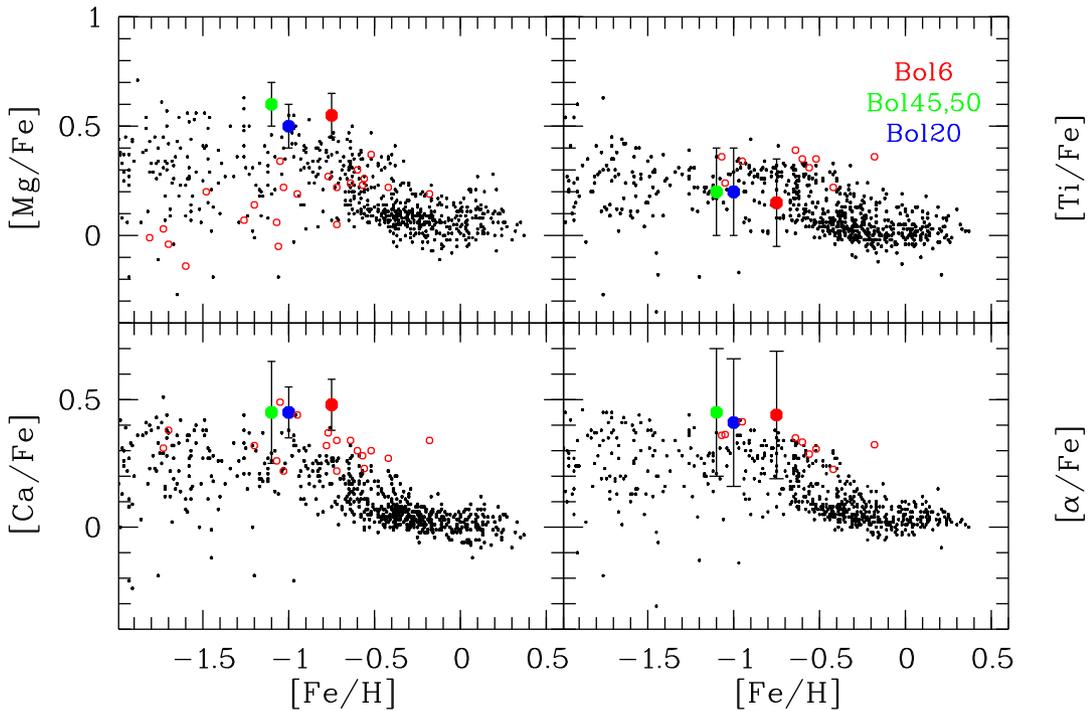} 
\caption{The same as in Fig.\ref{fig3}, but the data for GCs in M31 were determined by S16.}
\label{fig4}
\end{figure*}

\begin{figure*}
\includegraphics[scale=0.55, angle=270]{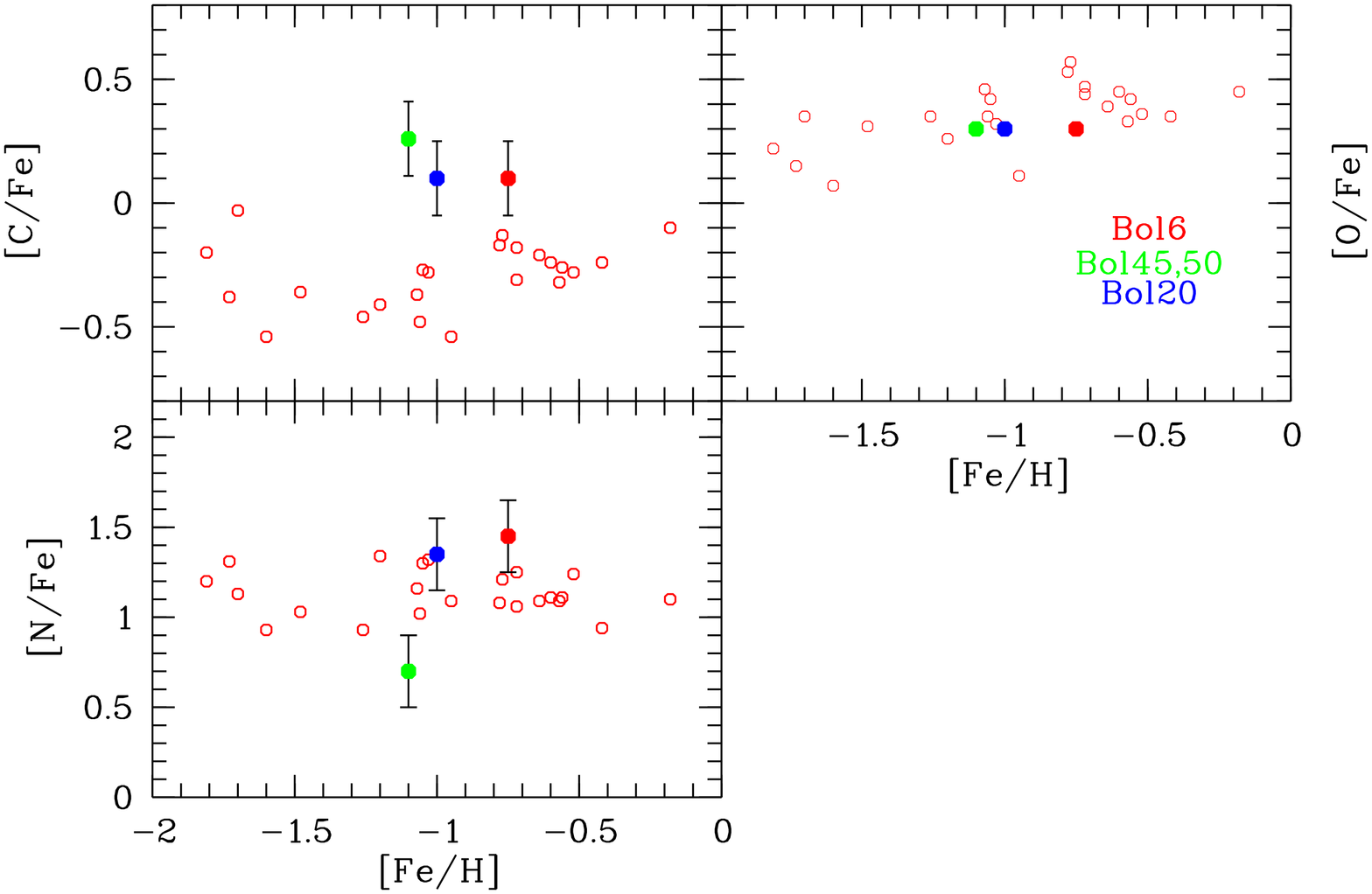} 
\caption{C, N and O abundances measured by MMI21 for Bol~6, Bol~20, Bol~45, and Bol~50 (filled circles) in  comparison with that of GCs in M31 estimated by S16.\label{fig5}}
\end{figure*}

\begin{figure*}
\includegraphics[scale=0.55, angle=270]{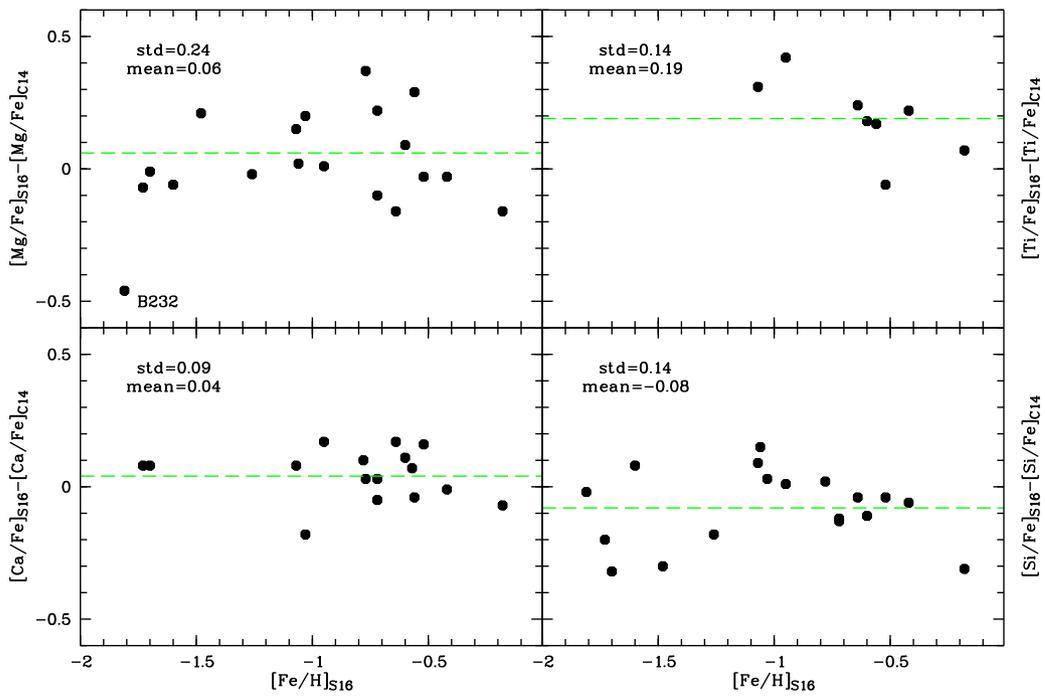} 
\caption{Comparison between the data from C14 and S16 for 19 common objects.}
\label{fig6}
\end{figure*}

\begin{figure*}
\includegraphics[scale=0.3, angle=180]{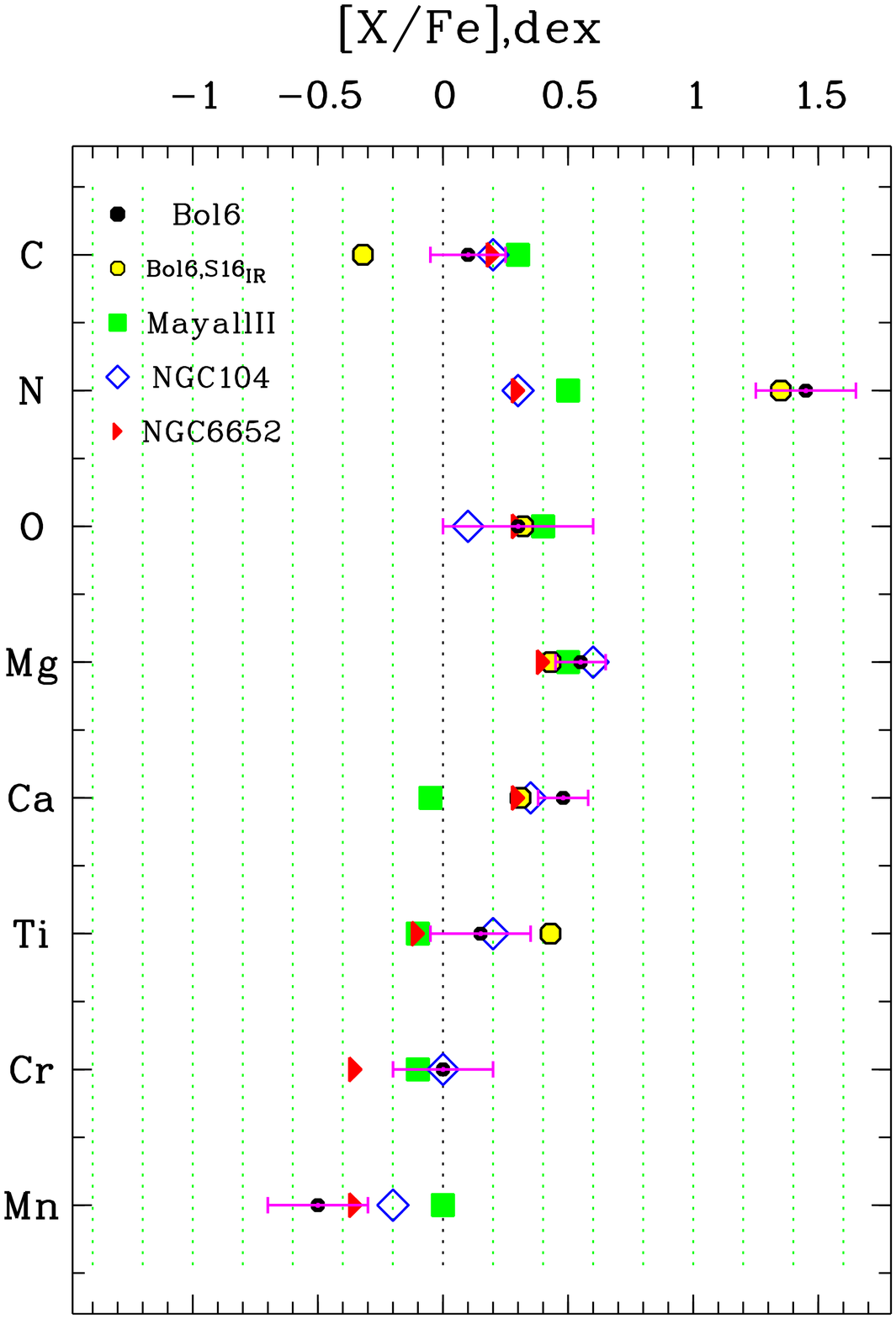} 
\includegraphics[scale=0.3, angle=180]{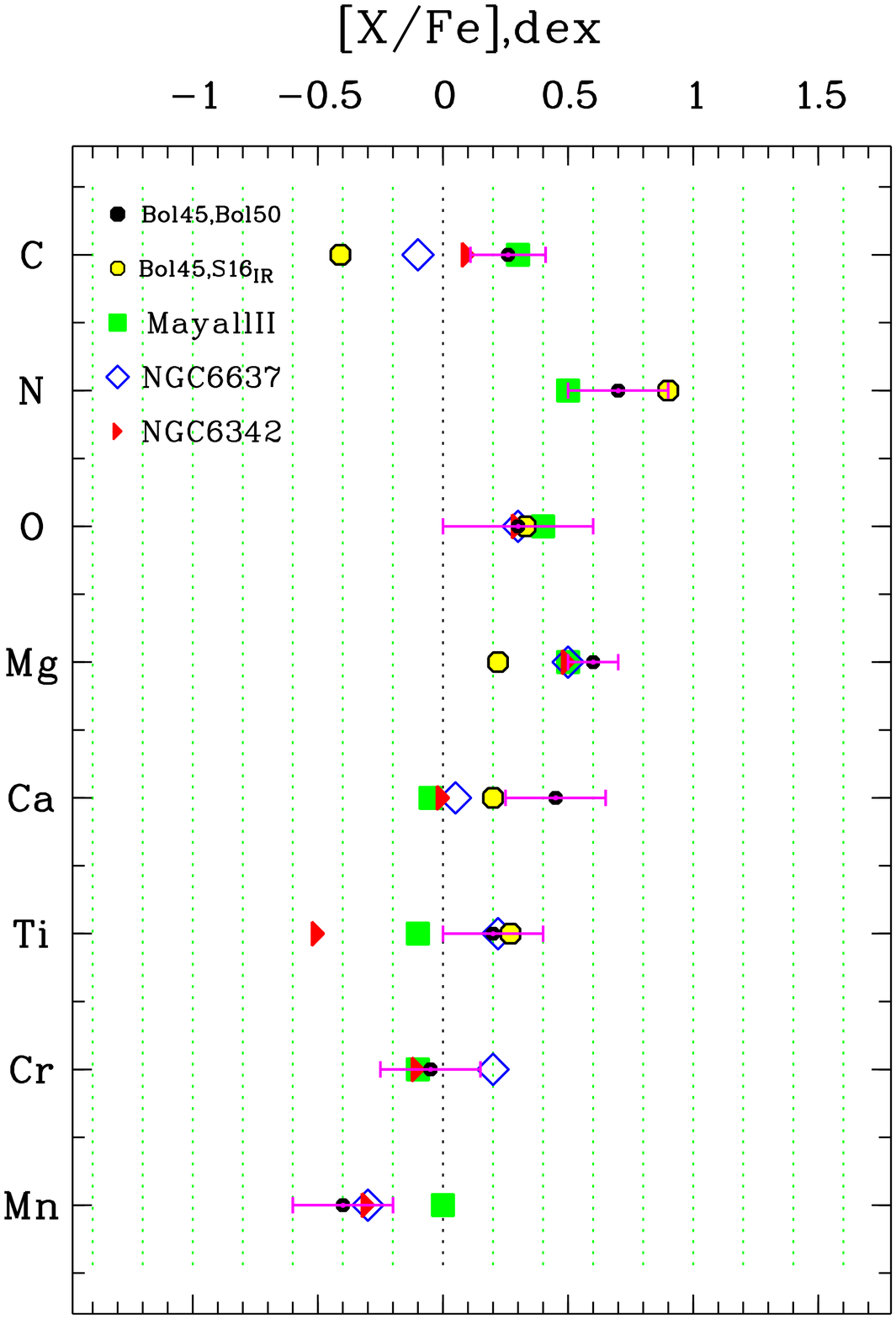}
\includegraphics[scale=0.3, angle=180]{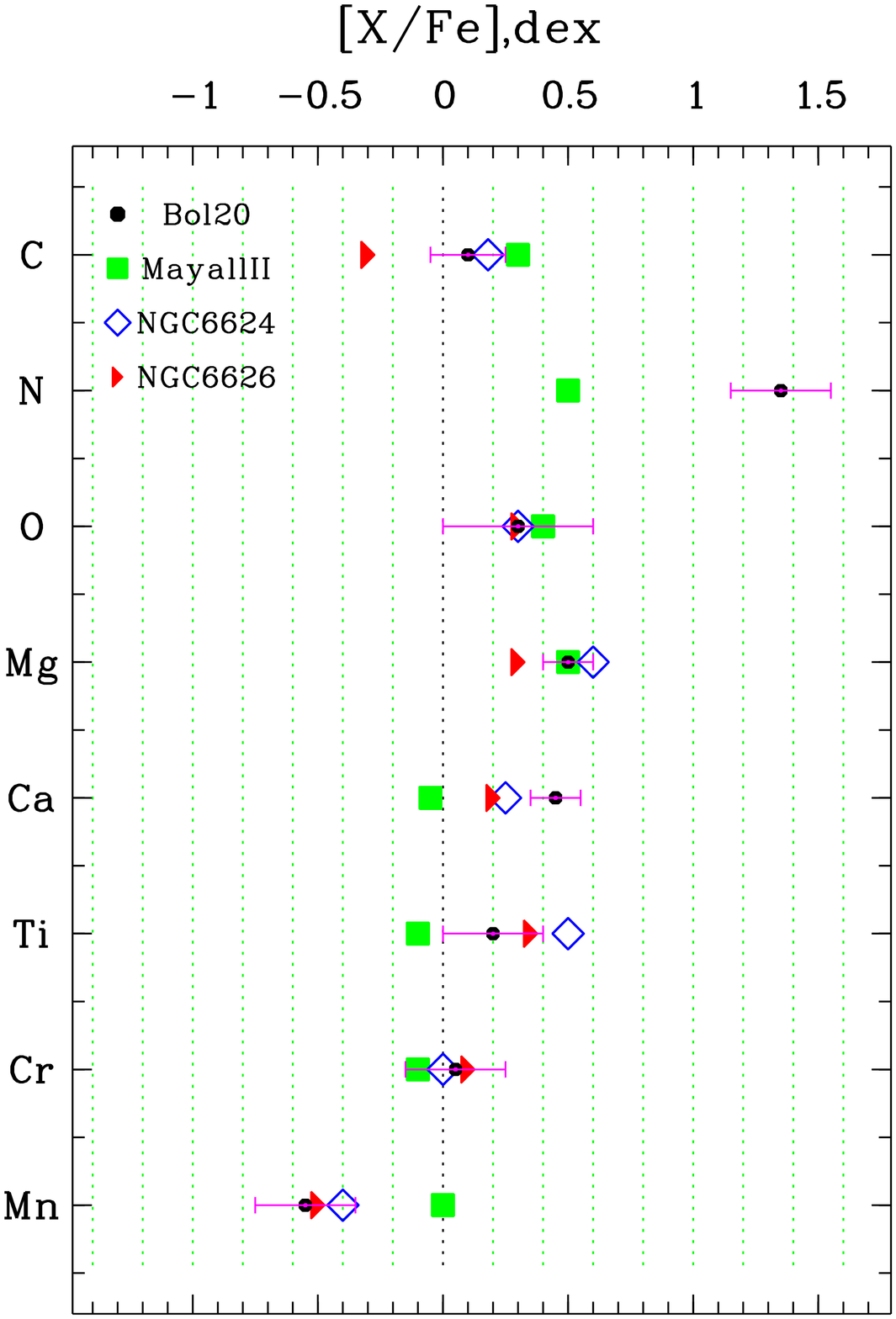}
\caption{Abundances of chemical elements determined by MMI21 for Bol~6 (left panel), Bol~45 and Bol~50 (middle panel), and Bol~20 (right panel) using the method of \cite{Sh20} (black dots) in comparison with the high-resolution integrated-light spectroscopic abundances from S16 (open circles) and with the abundances for Galactic GCs from Sharina et al.(2018, 2020) (symbols defined in the up left corners of the plots). Note that MMI21 derived the identical abundances for Bol~45 and Bol~50.}
\label{fig7}
\end{figure*}

In Fig.~\ref{fig3}, we compare [Mg/Fe], [Ca/Fe], and [Ti/Fe] abundances for Bol~6, Bol~20, Bol~45, and Bol~50 (MMI21) with the corresponding abundances for GCs
 in M31 from C14 (red circles). Black points represent the data for Galactic field stars from \cite{Vnn04}. In Fig.~\ref{fig4}, the same abundances for the 
Galactic stars are accomplished by the abundances for GCs in M31 from S16 (red circles). Figs.~\ref{fig3} and \ref{fig4} represent also the abundances of
 $\rm \alpha$-process elements from C14 and S16 for the GCs and stars. Alpha-element ratios of GCs were computed by C14 and S16 as follows:
 $\rm [\alpha/Fe] = ([Si/Fe]+[Ca/Fe]+[Ti/Fe])/3$. \cite{Vnn04} averaged Mg, Ca, and Ti abundances to calculate [$\rm \alpha$/Fe]. MMI21 computed [$\rm \alpha$/Fe]
 as the mean of Mg, Ca, and O abundances. Fig.~\ref{fig5} represents C, N, and O abundances from S16 in comparison with the abundances for our sample four GCs
 from MMI21. In Fig.~\ref{fig6}, we depict the differences in Mg, Ca, Ti, and Mg abundances between the data of S16 and C14 for 19 common objects in these two 
samples. Note that Figs.~\ref{fig3}-\ref{fig5} do not contain the data for Bol6 and Bol45 from C14 and S16 to avoid overlapping symbols. The data for Bol6 and 
Bol45 from S16 are presented in Fig.~\ref{fig7} together with the abundances for Bol6, Bol20, Bol45, and Bol50 from MMI21 and for Galactic GCs from \cite{Sh18}
 and \cite{Sh20}, who analysed IL spectra from \cite{Sch05}. The data from MMI21 are shown as black dots. The data from S16 for the same GCs are shown as open 
circles. Green filled squares demonstrate the data for Mayall~II in M31 from \cite{Sh18}. The \cite{Gal07} catalogue names of the GCs from C14 and S16 with 
$ \rm -1.1<[Fe/H] <-0.75$~dex are as follows: Bol34, 48, 63, 182, 235, 312, 381, 383, 386 and 403. Their ages estimated by C14 and \cite{Ca16} are older than 10 Gyr. 
Similar ages were found by MMI21 for Bol~6, Bol~20, Bol~45, and Bol~50. These GCs are located at a distance from the centre of M31 in projection on the sky:
 $ \rm 4.4 < R_{M 31} < 7.3$~kpc. The GCs with $ \rm -1.1<[Fe/H] <-0.75$~dex from the papers by C14 and S16 are distributed in projection on the sky in the wider
 high stellar density area around M31. Two of them (excluding two common objects) are pretty close to the GCs from MMI21: Bol~48 and Bol~63. Mg and Ca abundances
 are high for Bol~63, according to S16: $\rm [Mg/Fe]= 0.34\pm0.08, [Ca/Fe] = 0.49 \pm 0.10$. 

\begin{figure*}
\includegraphics[scale=0.55, angle=270]{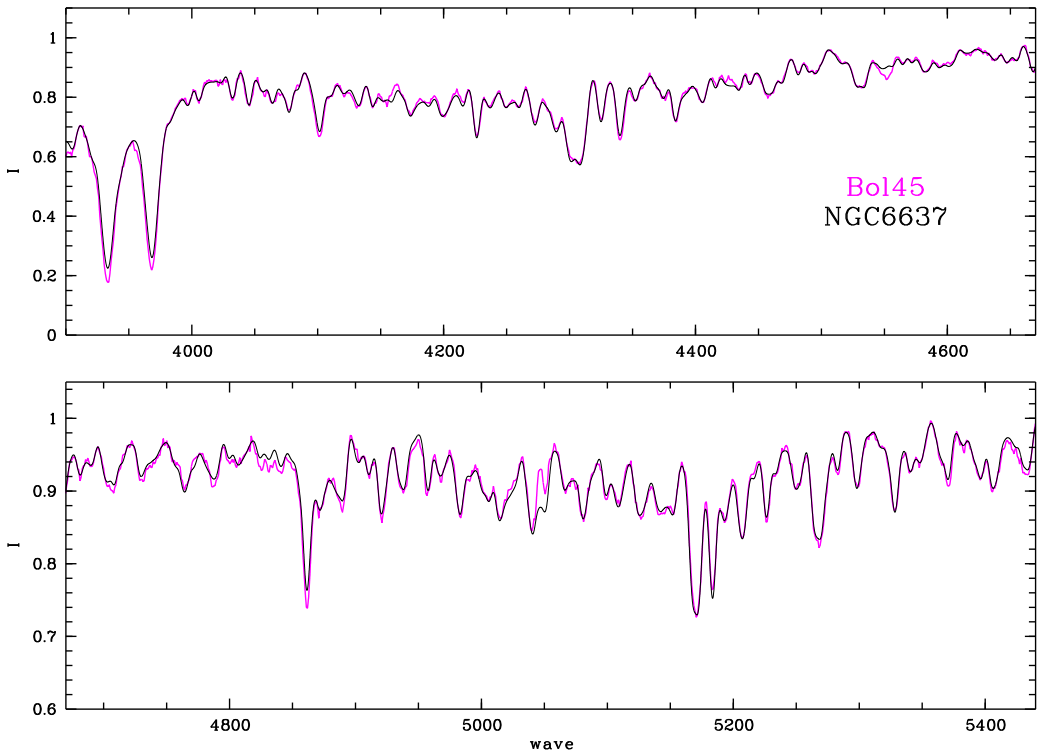} 
\caption{The IL spectrum of Bol45 from MMI21 (magenta) in comparison with the spectrum of Galactic GC NGC~6637 (\citealt{Sch05}) (black).}
\label{fig8}
\end{figure*}

After the inspection of Figs.~\ref{fig3}-\ref{fig7} one can make the following conclusions. (1) The enrichment with the $\rm \alpha$-process elements is about 
0.2 dex higher in four GCs studied by MMI21 than on average in Galactic field stars with $ \rm -1.1<[Fe/H] <-0.75$~dex (\citealt{Vnn04}). (2) The Mg, Ca, and Si
 abundances  for 19 GCs in S16 and C14 agree within the errors (Fig.\ref{fig6}). However, the objects in S16 appear to be systematically richer in Ti, especially 
at low metallicities. (3) We cannot find complete analogues for the studied clusters in our Galaxy and M31 in terms of age, helium content, and chemical 
composition. The closest analogue for Bol45 is the Galactic GC NGC6637 . The chemical composition of Bol~45 (MMI21) and NGC6637 (\citealt{Sh20}) is compared in
 Fig.~\ref{fig7}. The CMDs of the two GCs are compared in Fig.~\ref{fig1} . The comparison of their spectra can be seen in Fig.~\ref{fig8}. The shallower hydrogen
 lines in the spectrum of NGC6637 can be explained by the lower helium content. \cite{Sh20} selected the following isochrone (\citealt{B08}) to approximate the 
spectrum of NGC6637 by the synthetic one: Z=0.002, Y=0.23 and T=11.2~Gyr. These parameters are close to the ones  defined by MMI21 for Bol~45, with the exception
 of Y (Fig.~\ref{fig1}, panel c). (4) The chemical abundances of Bol6, Bol20, Bol45, and Bol50 measured by MMI21 agree with that in S16 and C14 (Fig.~\ref{fig7},
 see also Table~5 in MMI21 for the detailed comparison of the estimated abundances and age with literature values). The most significant exclusion is [C/Fe]. (5) 
[C/Fe] values of GCs with the metallicities $ \rm -1.1<[Fe/H] <-0.75$~dex in S16 are systematically lower than the C abundances of four GCs in MMI21 
(Fig.~\ref{fig5}). The reason was discussed by S16. The infrared spectral range used by these authors is mainly sensitive to the radiation of bright red giants,
 characterized by lower [C/Fe] than in the case of stars at earlier evolutionary stages. On the other hand, IL spectra in the optical wavelength range are 
sensitive more to the radiation of Main sequence stars.The systematic shifts between the Ca and Ti abundances at $ \rm -1.1<[Fe/H] <-0.75$~dex measured by MMI21, 
S16, and C14 should be attributed mainly to the differences in the applied methods. 

The presence of multiple stellar populations in GCs can reduce Mg and increase helium mass fraction, N and Ca abundances (\cite{CB21} and references therein). 
We cannot judge unambiguously about the presence of multiple populations in  Bol~6, Bol~20, Bol~45, and Bol~50 using the data from MMI21. MMI21 concluded that
 the obtained abundances correspond to those in the models of the chemical evolution of the Galaxy under the influence of supernovae type II (SNeII) and 
hypernovae (\citealt{Kob06}) in the metallicity range [Fe/H] = -1.1 ..- 0.75 dex. MMI21 discovered that the metallicity of four GCs is lower than the average 
metallicity of red giants in the M31 halo at a given distance from the centre of M31 (\cite{Gil20} and references therein). Using the results of \cite{Gil20}, 
MMI21 concluded that the average abundance of alpha elements in the stars of the inner halo of M31 ($ \rm [\alpha/Fe] = 0.45 \pm 0.09$~dex) is higher than in 
the stars of the outer halo ($ \rm [\alpha/Fe] = 0.3 \pm 0.16$~dex). The obtained [$\rm \alpha$/Fe] values for four GCs correspond to the average 
[$\rm \alpha$/Fe] value for stars in the inner halo at a given distance from the centre of M31. 

\section{Conclusion}

In this article we compared  the results of MMI21 on the age, Y, and abundances of Fe, C, N, Mg, Ca and Ti determination for four GCs in M31 with the available 
literature data for stars and GCs in two galaxies. We ascertained that four GCs have higher alpha-element abundances at their metallicities than the majority 
of GCs and field stars in M31 and our Galaxy. MMI21 noted that these alpha-element abundances can be described by the models of the chemical evolution under 
the influence of SNeII and hypernovae (\citealt{Kob06}).

According to MMI21, the obtained abundances of alpha-process elements in four clusters correspond to the average value for stars in the M31 inner halo at a 
given distance from the centre of M31. The metallicity of the studied clusters is lower than the average metallicity of red giants in the M31 halo 
(\citealt{Gil20}).  All these facts likely indicate that violent star forming events at the time of the GC formation lead to the high influence of SNeII on
 their chemical composition. The relatively low metallicities of four GCs can be the indication of the inflows of fresh intergalactic gas.

Further studies of ages and chemical composition of GCs and stars in M31 and its dwarf satellites will help to reveal the origin of GCs and the 
galactic subsystems they belong to.

\section{Acknowledgements}

We thank the organizing committee of the annual conference “Modern stellar astronomy” held in Moscow in 2021 in the Sternberg Astronomical
 Institute for the possibility to discuss these results.

Funding information: The authors state no funding involved.

Author contributions: All authors have accepted responsibility for the entire content of this manuscript and approved its submission.

Conflict of interest: Authors state no conflict of interest.

Data availability statement: The data underlying this article are available in the article.

%\section{Acknowledgements}
%The authors are grateful to V.V.~Shimansky. for providing a modified version of the program for modelling of IL spectra.
%This reasoning is partially valid for the explaining of lower Mg abundances in S16 for GCs with [Fe/H]$\sim -1$~dex in comparison with that in MMI21.



\end{document}